\documentclass[twocolumn,showpacs,preprintnumbers,amsmath,amssymb]{revtex4}
\usepackage{pifont}

\usepackage{graphicx}
\usepackage{dcolumn}
\usepackage{bm}

\begin{document}

\preprint{APS/123-QED}

\title{Ergodic Properties of Fractional Brownian-Langevin Motion
}

\author{Weihua Deng$^{1,2}$}

\author{Eli Barkai$^1$}%

\affiliation{%
$^1$Department of Physics, Bar Ilan University, Ramat-Gan 52900,
Israel
\\
$^2$School of Mathematics and Statistics, Lanzhou University,
Lanzhou 730000, PR China }%



\begin{abstract}
We investigate the time average mean square displacement
$\overline{\delta^2}(x(t))=\int_0^{t-\Delta}[x(t^\prime+\Delta)-x(t^\prime)]^2
dt^\prime/(t-\Delta)$ for fractional Brownian and Langevin motion.
Unlike the previously investigated  continuous time random walk
model $\overline{\delta^2}$ converges to the ensemble average
$\langle x^2 \rangle \sim t^{2 H}$ in the long measurement time
limit. The convergence to ergodic behavior is however slow, and
surprisingly the Hurst exponent $H=3/4$ marks the critical point of
the speed of convergence. When $H<3/4$, the ergodicity breaking
parameter $\mbox{EB} = \mbox{Var} ( \overline{\delta^2} ) / \langle
\overline{\delta^2} \rangle^2\sim  k(H) \cdot\Delta\cdot t^{-1}$,
when $H=3/4$, $\mbox{EB} \sim (9/16)(\ln t) \cdot\Delta \cdot
t^{-1}$, and when $3/4<H <1, \mbox{EB} \sim k(H)\Delta^{4-4H}
t^{4H-4}$. In the ballistic limit $H \to 1$ ergodicity is broken and
$\mbox{EB} \sim 2$. The critical point $H=3/4$ is marked by the
divergence of the coefficient $k(H)$. Fractional Brownian motion as
a model for recent experiments of sub-diffusion of mRNA in the cell
is briefly discussed and comparison with the continuous time random
walk model is made.
\end{abstract}

\maketitle

\section{Introduction}

Fractional calculus, e.g. $d^{1/2}/ dt^{1/2}$,
 is a powerful mathematical tool for the investigation of
physical and biological phenomena with long-range correlations or
long memory \cite{Metzler:00}. For example fractional calculus
describes the mechanical memory of viscoelastic materials
\cite{Mainardi:88}. An important application of fractional calculus
is in the stochastic modeling of anomalous diffusion. Fractional
Fokker-Planck equations describe the long time behavior of the
continuous time random walk (CTRW) model, when waiting times  and/or jump
lengths have power-law distributions
\cite{Barkai:00,Barkai:01,Metzler:00,Deng:07}. A different
stochastic approach to anomalous diffusion is based on fractional
Brownian motion (fBM) \cite{Mandelbrot:68} which is related to
recently investigated fractional Langevin equations (see details
below) \cite{Goychuk:07,Min:05,Burov:08}.

Recent single particle tracking  of mRNA molecules \cite{Golding:06}
and lipid granules \cite{Tolic:04} in living cells revealed that
time averaged mean square displacement $\overline{\delta^2}$
(defined below more precisely) of individual particles remains a
random variable while indicating that the particle motion is
sub-diffusive.  This means that the time averages are not identical
to ensemble averages. Such breaking of ergodicity was investigated
within the sub-diffusive CTRW model \cite{He:08,Klafter:08}. It was
shown that transport and diffusion constants extracted from single
particle trajectories remain random variables, even in the long
measurement time limit. For a non-technical point of view on this
problem see \cite{Sokolov:08}. Here we consider three stochastic
models of anomalous diffusion:  fBM  and the under-damped and the
over-damped fractional Langevin equation. Except for limiting cases
(i.e. ballistic diffusion) we find that the time average
$\overline{\delta^2}$, in the long measurement time limit, is
identical to the ensemble average $\langle x^2 \rangle$, indicating
that these models are ergodic. Note however, that experiments  on
anomalous dynamics of particles in the cell are always conducted for
finite times (due to the life time of the cell).  Here we find the
finite time corrections to ergodic behavior, namely we give
estimates on how far will a finite time measurement of anomalous
diffusion deviate from the ensemble average. Since the convergence
to ergodic behavior is slow our results seem particularly important
to finite time experiments. The problem of estimating diffusion
constants from single particle tracking, for normal diffusion,  is
already well investigated \cite{Saxton:97}.

 In recent years there
was much interest in non-ergodicity of anomalous diffusion
processes. A well investigated system are blinking quantum dots
\cite{Dahan:03,Margolin:05}, which exhibit a L\'evy walk type of
dynamics (a super-diffusive process). Very general formula for the
distribution of time averages for weakly non ergodic systems was
derived in \cite{Rebenshtok:07}, and this framework was shown to
describe the sub-diffusive CTRW \cite{Bel:05}.  Bao et al have
investigated ergodicity breaking for stochastic dynamics described
by the generalized Langevin equation \cite{Bao:05}. They considered
the time averaged velocity variance. The latter converges to $k_B T
/ m$ in thermal equilibrium if the process is ergodic. It was shown
\cite{Bao:05} that under certain conditions the generalized Langevin
equation is non ergodic (see also
\cite{Lee:01,Oliv:03,Costa:06,Dhar:07,Plyukhin:08}). Our work,
following the recent experiments \cite{Tolic:04,Golding:06},
considers the time average of the mean square displacements which
yields information on diffusion constants. In contrast with
\cite{Bao:05}, we consider an out of equilibrium situation, because
our observable: the coordinate of the particle does not reach an
equilibrium, since the system is infinite.

\section{Stochastic Models}

\subsection{Fractional Brownian Motion}

 Fractional Brownian motion is
generated from fractional Gaussian noise, like Brownian motion from
white noise. Mandelbrot and van Ness \cite{Mandelbrot:68} defined
fBM with
\begin{equation}
\begin{array}{c}
\displaystyle B_H(t):=\frac{1}{\Gamma(H+\frac{1}{2})}\left( \int_0^t
(t-\tau)^{H-\frac{1}{2}}d
B(\tau)\right.\\
\displaystyle
\left.+\int_{-\infty}^0[(t-\tau)^{H-\frac{1}{2}}-(-\tau)^{H-\frac{1}{2}}]
d B(\tau) \right),
\end{array} \label{eq:one}
\end{equation}
where $\Gamma$ represents the Gamma function and $0<H < 1$ is called
the Hurst parameter. The integrator $B$ is ordinary Brownian motion.
Note that $B$ is recovered when taking $H=1/2$. The right hand side
of Eq. (\ref{eq:one}) is the sum of two independent Gaussian
processes. In the definition, for the first Gaussian process, we
identify the so-called fBM of Riemann-Liouville type \cite{Lim:02}.
Standard fBM, i.e. Eq. (\ref{eq:one}),
is the only Gaussian self-similar process with stationary increments
\cite{Mandelbrot:68}. The variance of $B_H(t)$ is $2D_H t^{2H}$,
where $D_H=(\Gamma(1-2H)\cos(H\pi))/(2H\pi)$. In the following, for
some given $H$, we denote the trajectory sample of fBM
 $x(t)$. The properties that uniquely characterize the
fBM can be summarized as follows: $x(t)$ has
stationary increments; $x(0)=0$ and $\langle x(t) \rangle=0$ for $t
\ge 0$; $\langle x^2(t) \rangle=2D_H t^{2H}$ for $t \ge 0$; $x(t)$
has a Gaussian distribution for $t>0$. From the above properties,
the covariance function is \cite{Samorodnitsky:94},
\begin{equation} \langle x(t_1) x(t_2)
\rangle=D_H(t_1^{2H}+t_2^{2H}-|t_1-t_2|^{2H} ),~~t_1,t_2>0.
\label{eq:two}
\end{equation}
The non-independent  increment process of fBM, called fractional
Gaussian noise (fGn), is given by
\begin{equation}
\xi(t)=\frac{dx(t)}{dt},~~ t>0, \label{eq:three}
\end{equation}
which is a stationary Gaussian process and has a standard normal
distribution for any $t>0$.
The mean $\langle \xi(t) \rangle =0$ and
the covariance is
\begin{equation}
\langle \xi(t_1) \xi(t_2) \rangle
 = 2D_H H(2H-1)|t_1-t_2|^{2H-2},~~
t_1,t_2>0. \label{eq:four}
\end{equation}

\subsection{Fractional Langevin Equation}

The standard Langevin equation with white noise can be extended to
a generalized Langevin equation with a power-law memory kernel. Such
an approach was recently used to model dynamics of single proteins
by the group of Xie  \cite{Min:05} and can be derived from the
Kac-Zwanzig model of Brownian motion \cite{Kupferman:04}. The
under-damped fractional Langevin equation reads
\begin{equation}
m \frac{d^2y(t)}{dt^2}=-\bar{\gamma} \int_0^t(t-\tau)^{2H-2}
\frac{dy}{d\tau} d\tau +\eta\cdot\xi(t),
 \label{eq:five}
\end{equation}
where according to the fluctuation dissipation theorem
$$\eta=\sqrt{\frac{k_BT\bar{\gamma}}{2D_HH(2H-1)}},$$
 $\xi(t)$
is fGn defined in Eqs. (\ref{eq:three}) and (\ref{eq:four}),
$1/2<H<1$ is the Hurst parameter, $\bar{\gamma}>0$ is a generalized
friction constant. Eq. (5) is called a fractional Langevin equation
since the memory kernel yields a fractional derivative of the
velocity (hint use Laplace transform or see
\cite{Metzler:00,Li:07}). Note that if $0<H<1/2$ the integral over
the memory kernel diverges, and hence it is assumed that $1/2<H<1$.
This leads to sub-diffusive behavior $\langle y^2 \rangle \sim
t^{2-2H} $. An over-damped fractional Langevin equation, where
Newton's acceleration term is neglected reads \cite{Min:05}
\begin{equation}
0=-\bar{\gamma} \int_0^t(t-\tau)^{2H-2} \frac{dz}{d\tau} d\tau
+\eta\cdot\xi(t).
 \label{eq:six}
\end{equation}
In what follows we investigate ergodic properties of the processes
$x(t),y(t)$ and $z(t)$.

\section{Ergodic Properties}

We consider the time average
\begin{equation}
\overline{\delta^2}(x(t))=\frac{\int_0^{t-\Delta}[x(t^\prime+\Delta)-x(t^\prime)]^2
dt^\prime}{t-\Delta},\label{eq:seven}
\end{equation}
which is a random variable depending on the stochastic
path $x(t)$.  Here $\Delta$ is called the lag time. As is
well-known, for normal Brownian motion, the ensemble average mean
square displacement is $\langle x^2(t) \rangle=2 D_1 t$ while  the
time average mean square displacement of the single trajectory
$\overline{\delta^2}(x(t))=2 D_1 \Delta$ in statistical sense and in
the long measurement time limit. Hence we may use in experiment a
single trajectory of a Brownian particle to estimate the diffusion
constant $D_1$. Will similar ergodic behavior be found also for
fBM?

\subsection{Fractional Brownian Motion}
If the average of  the random variable $\overline{\delta^2}$ is
equal to the ensemble average $\langle x^2 \rangle$, and if the
variance of $\overline{\delta^2}$ tends to zero when the measurement
time is long the process is ergodic, since then the distribution of
$\overline{\delta^2}$ tends to a delta function centered on the
ensemble average. For fBM, using Eqs. (\ref{eq:two}) and
(\ref{eq:seven})

\begin{equation}
\langle\overline{\delta^2}(x(t))\rangle=\frac{\int_0^{t-\Delta}\langle[x(t^\prime+\Delta)-x(t^\prime)]^2\rangle
\textrm{d}t^\prime}{t-\Delta}=2D_H\Delta^{2H},\label{eq:eight}
\end{equation} hence $\langle \overline{\delta^2} \rangle = \langle
x^2 \rangle$ for all times. The variance of
$\overline{\delta^2}(x(t))$ is,

\begin{equation} {\rm Var}(\overline{\delta^2}(x(t)))=\left\langle
\left(\overline{\delta^2}(x(t))\right)^2\right\rangle-\left\langle\overline{\delta^2}(x(t))\right\rangle^2.
\label{eq:nine} \end{equation}

\begin{figure}[!htb]
\begin{center}
\includegraphics[width=7.5cm,height=5cm]{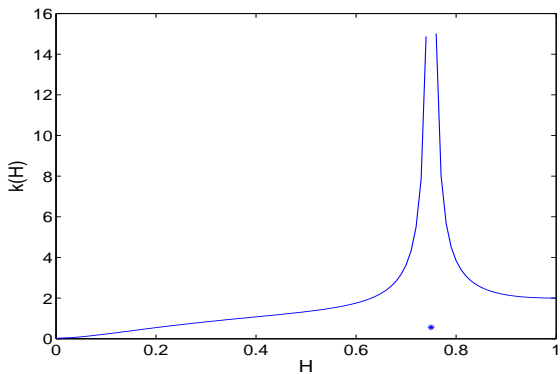}
\caption{\label{fig1} The function of $k(H)$ Eq. (\ref{eq:Coeff}).}
\end{center}
\end{figure}

\begin{figure}[!htb]
\begin{center}
\includegraphics[width=9cm,height=6cm]{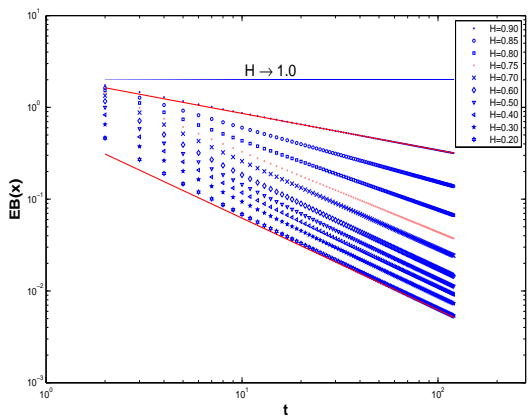}
 \caption{\label{fig2} The EB parameter Eq.
(\ref{eq:MainResult}) for fractional Brownian motion $x(t)$ versus
time $t$, for different values of Hurst exponent. Here we present
exact results obtained  directly by calculating Eqs.
(\ref{eq:eight},\ref{eq:twelve},\ref{eq:thirteen}). The two solid
lines (red on line) are the asymptotic theory Eq.
(\ref{eq:MainResult}) for $H=0.2$ and $H=0.9$. In the ballistic
limit $H \to 1$ we get non-ergodic behavior. Notice that for
$H<3/4$, and long $t$ the curves are parallel to each other due to
the  ${\rm EB} \sim t^{-1} $ law valid for $H<3/4$, while for
$H>3/4$ the slopes are changing as we vary $H$. The lag time is
$\Delta=1$. }
\end{center}
\end{figure}

\begin{figure}[!htb]
\begin{center}
\includegraphics[width=9.2cm,height=6cm]{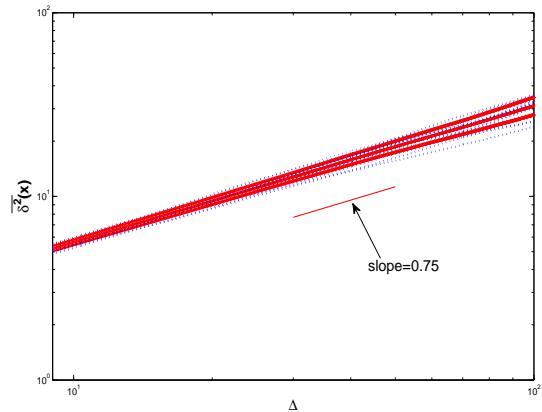}
\caption{\label{fig3} We simulate fBM and present the time average
$\overline{\delta^2}(x(t))$ versus $\Delta$ (dotted curves blue on
line). We show  $23$ trajectories, the solid line in the middle
being the average of the trajectories. We observe
$\overline{\delta^2}(x(t))\propto \Delta^{3/4}$, similar to that in
\cite{Tolic:04,Golding:06}. The measurement time is
$t=10^4,~H=3/8,~D_H=1/2$. Line with a slope of $0.75$ is drawn to
guide the eye.
We also show $\langle \overline{\delta}^2 \rangle \pm
\sqrt{\mbox{Var}(\overline{\delta^2}) }$ (two solid lines red on
line) obtained from Eqs. (\ref{eq:eight}, \ref{twenty-three}) which
give an analytical estimate on the scatter of the data. }
\end{center}
\end{figure}

A dimensionless measure of ergodicity breaking (EB) is the
parameter,
\begin{equation} {\rm EB}(x(t))=\frac{{\rm
Var}(\overline{\delta^2}(x(t)))}{\left\langle
\overline{\delta^2}(x(t)) \right\rangle^2}, \label{eq:EB}
\end{equation}
which is zero in the limit $t \to \infty$ if the process is ergodic.
In the next sub-section we derive our main result, valid for large
$t$, we find:
\begin{equation} {\rm EB}(x(t)) \thicksim \left\{
\begin{array}{ll}
k(H)
\frac{\Delta}{t},& 0<H<\frac{3}{4}, \\
\\
k(H) \frac{\Delta}{t}\ln t,& H=\frac{3}{4},
\\
\\
k(H)(\frac{\Delta}{t})^{4-4H},& \frac{3}{4}<H<1,
\end{array}
\right. \label{eq:MainResult}
\end{equation}
\begin{widetext}
where
\begin{equation}
 k(H) =\left\{
 \begin{array}{ll}
  \int_0^{\infty}((\tau+1)^{2H}+|\tau-1|^{2H}-2\tau^{2H})^2d
 \tau, & 0<H<\frac{3}{4}, \\
\\
4H^2(2H-1)^2=\frac{9}{16}, & H=\frac{3}{4}, \\
\\
(\frac{4}{4H-3}-\frac{4}{4H-2})H^2(2H-1)^2, & \frac{3}{4}<H<1.
\end{array}
\right. \label{eq:Coeff}
\end{equation}
\end{widetext}
The most remarkable result is that $k(H)$ diverges when $H\to 3/4$,
$H=3/4$ marks a non smooth transition in the properties of
fractional Brownian motion, see Fig. \ref{fig1}. Notice that $k(H)
\to 0$ when $H \to 0$ so the asymptotic convergence is expected to
hold only after very long times when $H$ is small, since then the
diffusion process is very slow. When $H \to 1$ we have ${\rm
EB}(x(t))\sim 2 $ indicating ergodicity breaking, see Fig.
\ref{fig2}.

Fig. \ref{fig3} displays the simulations of
$\overline{\delta^2}(x(t))$ showing the randomness of
 the time average for finite time measurements.
In this simulation we generate single trajectories using the Hosking
method \cite{Hosking:84}, and then perform the time average to find
$\overline{\delta^2}$. The Fig. mimics the experimental results on
single lipid granule in a yeast cell and of mRNA molecules inside a
living E-coli cells \cite{Tolic:04,Golding:06}, where $H=3/8$ was
recorded. Note however that the scatter of the experiments data
seems larger (see Figs. in \cite{Tolic:04,Golding:06}), at least
with the naked eye. Further we did not consider in our simulations
the effect of the cell boundary. Direct comparison at this stage
between experiments and stochastic theory is impossible, since the
number of measured trajectories is small.

\begin{widetext}
\subsection{Derivation of Main Result Eq.(\ref{eq:MainResult})}

From Eq. (\ref{eq:eight}),
\begin{equation}
\left\langle \left(\overline{\delta^2}(x(t))\right)^2\right\rangle =
\frac{\int_0^{t-\Delta}\textrm{d}t_1 \int_0^{t-\Delta}\textrm{d}t_2
\langle [x(t_1+\Delta)-x(t_1)]^2[x(t_2+\Delta)-x(t_2)]^2
\rangle}{(t-\Delta)^2}. \label{eq:ten}
\end{equation}
Using Eq. (\ref{eq:two}) and the following formula for Gaussian
process with mean zero \cite{Kubo:95},
$$
\langle x(t_1)x(t_2)x(t_3)x(t_4) \rangle=\langle
x(t_1)x(t_2)\rangle\langle x(t_3)x(t_4) \rangle+\langle
x(t_1)x(t_3)\rangle\langle x(t_2)x(t_4) \rangle+\langle
x(t_1)x(t_4)\rangle\langle x(t_2)x(t_3) \rangle,
$$
we obtain
\begin{equation}
\langle [x(t_1+\Delta)-x(t_1)]^2[x(t_2+\Delta)-x(t_2)]^2 \rangle
=4D_H^2\Delta^{4H}+2D_H^2\{|t_1+\Delta-t_2|^{2H}+|t_2+\Delta-t_1|^{2H}-2|t_1-t_2|^{2H}
\}^2. \label{eq:eleven}
\end{equation}
From Eqs.
(\ref{eq:eight},\ref{eq:nine},\ref{eq:ten},\ref{eq:eleven}), we have
\begin{eqnarray}
{\rm Var}(\overline{\delta^2}(x(t))) 
&=& \underbrace{
4D_H^2\left\{\int_0^{2\Delta}(t-\Delta-t^\prime)\{(t^\prime+\Delta)^{2H}+|t^\prime-\Delta|^{2H}-2
{t^\prime}^{2H}\}^2
\textrm{d}t^\prime\right\}/(t-\Delta)^2}_{V_1} \label{eq:twelve}\\
&& +
\underbrace{4D_H^2\left\{\int_{2\Delta}^{t-\Delta}(t-\Delta-t^\prime)\{(t^\prime+\Delta)^{2H}+(t^\prime-\Delta)^{2H}-2
{t^\prime}^{2H}\}^2 \textrm{d}t^\prime\right\}/(t-\Delta)^2}_{V_2}.
\label{eq:thirteen}
\end{eqnarray}
%
 When $t>>\Delta$ we may approximate the upper limit in the integral
  of $V_2$ with $t$, and $1/(t-\Delta)^2 \to 1/t$. We then make a change of
  variables according to $x=(t - t')/t$ and find
  \begin{equation}
  V_2 = 4 D_H^2 t^{4 H} \int_0 ^1 x (1 - x)^{4 H} \left[ \left( 1 + {\Delta\over t(1 - x)} \right)^{2 H} + \left|1 -
   {\Delta\over t(1 - x)}\right|^{2H}-2\right]^2{\rm d}x.
  \end{equation}
  We expand in $\Delta/t$ to second order
  and find
  \begin{equation}
  V_2 \sim 4 D_H^2 t^{4 H} \left( {\Delta \over t} \right)^4 H^2 \left( 2 H - 1 \right)^2 \int_0 ^1 x ( 1 - x)^{4 H -4} {\rm d} x.
  \end{equation}
 The integral is finite only if $H> 3/4$ hence for $H\leqslant3/4$
 we will soon use a different
 approach. We see that $V_2 \sim t^{4 H - 4}$ while it is easy to show that
 $V_1\sim 1/t$ hence for $H > 3/4$ we find after solving the integral
 \begin{equation}
{\rm Var} ( \overline{\delta^2}(x(t))) \sim 16 D_H^2 t^{4 H} \left(
{\Delta \over t } \right)^4 H^2 (2 H - 1)^2 \left( { 1 \over 4 H -
3} - { 1 \over 4 H- 2} \right). \label{nineteen}
 \end{equation}
 Now we write the variance as
\begin{equation}
{\rm Var} ( \overline{\delta^2}(x(t)))= { 4 D_H^2 \over (t -
\Delta)^2} \int_0^{t - \Delta} \left(t-\Delta-t^\prime \right)
\left[ \left( t^\prime+\Delta\right)^{2H}+|t^\prime-\Delta|^{2H}-2
{t^\prime}^{2H} \right]^2 {\rm d} t^\prime.
\end{equation}
Changing variables according to $\tau = t^\prime/\Delta$ we find
\begin{equation}
{\rm Var} ( \overline{\delta^2}(x(t))) = { 4 D_H^2 \over (t -
\Delta)} \Delta^{ 4 H + 1}  \int_0 ^{t/\Delta  - 1} {\rm d} \tau
\left[ \left( 1 + \tau\right)^{2 H} + | 1- \tau |^{ 2 H} - 2 \tau^{2
H} \right]^2 + {\rm Corr}. \label{twenty-one}
\end{equation}
The correction term is
\begin{equation}
\mbox{Corr} = - { 4 D_H^2 \over (t - \Delta)^2} \Delta^{4 H + 2}
\int_0 ^{ t/\Delta - 1} {\rm d} \tau  \left[ \left( 1 +
\tau\right)^{2 H} + |1- \tau|^{ 2 H} - 2 \tau^{2 H}
\right]^2\tau.\label{twenty-two}
\end{equation}
Taking the upper limit of the integral in Eq. (\ref{twenty-one}) to
$\infty$
we find that for $H<3/4$ and long times
\begin{equation}
{\rm Var} ( \overline{\delta^2}(x(t)))\sim  4 D_H^2 \Delta^{4 H}
\left( {\Delta \over t} \right)  \int_0 ^{\infty} {\rm d} \tau
\left[ \left( 1 + \tau\right)^{2 H} + | 1- \tau |^{ 2 H} - 2 \tau^{2
H} \right]^2. \label{twenty-three}
\end{equation}
This is because ${\rm Corr}\sim t^{\max\{4H-4,-2\}}$ (we prove this
in the following) and this term is smaller than the leading term
which has a $1/t$ decay, since $H< 3/4$.

Now we estimate the correction term Eq. (\ref{twenty-two})

\begin{equation} \frac{1}{t^2}\int_0 ^{ t/\Delta - 1} {\rm d} \tau
\left[ \left( 1 + \tau\right)^{2 H} + |1- \tau|^{ 2 H} - 2 \tau^{2
H} \right]^2\tau =\frac{1}{t^2}\left(\int_0^2 {\rm d} \tau+\int_2^{
t/\Delta - 1}{\rm d} \tau \right)  \left[ \left( 1 + \tau\right)^{2
H} + |1- \tau|^{ 2 H} - 2 \tau^{2 H} \right]^2\tau.
\end{equation}
Using the Lagrange reminder of Taylor expansion in $1/\tau$, when $H
\leqslant \frac{3}{4}$ we have
\begin{equation}
\begin{array}{l}
\displaystyle \frac{1}{t^2} \int_2^{ t/\Delta - 1}{\rm d} \tau
\left[ \left( 1 + \tau\right)^{2 H} + |1- \tau|^{ 2 H} - 2 \tau^{2
H} \right]^2\tau
\\
\\\displaystyle
= \frac{1}{t^2} \int_2^{ t/\Delta - 1}{\rm d} \tau   \left[ \left( 1
+ \frac{1}{\tau}\right)^{2 H} + \left|1 - \frac{1}{\tau}\right|^{ 2
H} - 2  \right]^2\tau^{4H+1}
\\
\\\displaystyle
=\frac{1}{t^2} \int_2^{ t/\Delta - 1}{\rm d} \tau \left[
(1+\xi)^{2H-2}+(1-\xi)^{2H-2} \right]2H(2H-1)\tau^{4H-3},~~~~ \xi
\in [0,\frac{1}{2}]
\\
\\\displaystyle
\sim t^{\max\{4H-4,-2\}}.
\end{array} \label{Lang}
\end{equation}
For $H=\frac{3}{4}$ we use Eq. (\ref{twenty-one}),  however now we
expand to third order and find

\begin{equation}
{\rm Var} ( \overline{\delta^2}(x(t)))\sim  16 D_H^2H^2(2H-1)^2
\Delta^3 \ln t \left( \frac{\Delta}{t} \right), \label{twenty-six}
\end{equation}
while the correction term ${\rm Corr} \sim t^{-1}$ [see Eq.
(\ref{Lang})] is negligible.  Using Eqs.
(\ref{nineteen},\ref{twenty-three},\ref{twenty-six}) we derive Eq.
(\ref{eq:MainResult}).

\end{widetext}

\subsection{Over Damped Fractional Langevin Equation }

We now analyze the over-damped fractional Langevin Eq.
(\ref{eq:six}), we can rewrite it in a convenient way as
\begin{equation}
\bar{\gamma}\Gamma(2H-1) D^{1-2H}D z(t)=\eta\cdot\xi(t),
 \label{eq:twenty-nine}
\end{equation}
where $D=d/dt$, and $D^{1-2H}$ is the Riemann-Liouville fractional
integral of $2H-1$ order. Using the tools of fractional calculus
\cite{Li:07}, we get

\begin{equation}
\bar{\gamma}\Gamma(2H-1)z(t)=\eta\cdot D^{2H-2}\xi(t).
 \label{eq:thirty}
\end{equation}
Then  since $\langle \xi(t) \rangle =0$ we have $\langle z(t)
\rangle=0$, and
\begin{equation} \langle z(t_1)z(t_2)
\rangle=D_F(t_1^{2-2H}+t_2^{2-2H}-|t_1-t_2|^{2-2H}),
 \label{eq:thirty-one}
\end{equation}
where
$$
D_F=\frac{k_B T\pi\csc[\pi
({2-2H})]}{\bar{\gamma}(2-2H)\Gamma^2(2H-1)\Gamma^2(2-2H)}.
$$
From Eq. (\ref{eq:thirty-one}) we learn that Eq.
(\ref{eq:twenty-nine}) exhibits the same behavior as fBM Eq.
(\ref{eq:one}) in the sub-diffusion case. Note that for fBM $\langle
x^2 \rangle \sim t^\alpha$ with $\alpha = 2 H$ while for the
fractional Langevin equation $\langle z^2 \rangle \sim t^\alpha$
with $\alpha = 2 - 2 H$, and of course
\begin{figure}[!htb]
\begin{center}
\includegraphics[width=9cm,height=8cm]{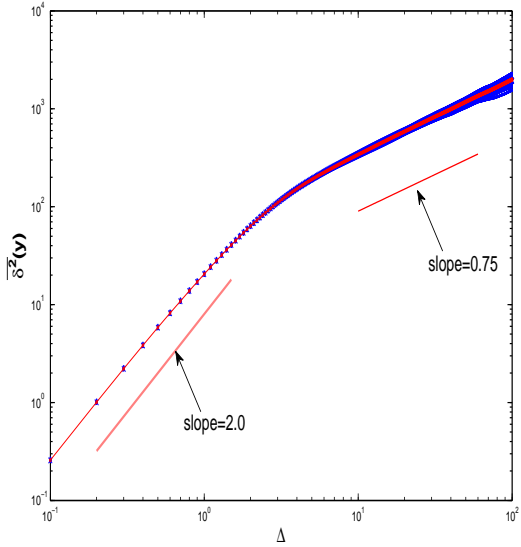}
\caption{\label{fig4} The time average $\overline{\delta^2}(y(t))$
 is a
random variable depending on the underlying trajectory. Total 23
trajectories, besides the solid line with mark $\circ$ being the
average of the 23 trajectories, are plotted. The measurement time is
$t=
10^4,2-2H=0.75,~D_H=1/2,~k_B=1,~m=1,~\bar{\gamma}=1,~v_0=1,~T=1$.
Lines with  slopes of $2.0$ (ballistic motion in short times) and
$0.75$ (sub-diffusion for long times) are drawn to guide the eye.
For long $\Delta$ the behavior of the under-damped motion is similar
to usual
fractional Brownian motion.} 
\end{center}
\end{figure}
the diffusion constants have different dependencies on parameters of
the noise. However these minor modifications do not change our main
result for $0<H<1/2$ obtained in the previous section (only switch
the value $2 H$ to $2-2H$). See this note that the $\mbox{EB}$
parameter depends on the behavior of correlation function Eq.
(\ref{eq:two}) and the latter are identical for the processes $x(t)$
and $z(t)$ in the sub-diffusion case, so ${\rm EB}(z) \sim {\rm EB}
(x)$.

\vskip 1cm

\subsection{Under Damped Fractional Langevin Equation}

We now analyze the fractional  Langevin equation with power-law
kernel, namely, Eq. (\ref{eq:five}),
\begin{equation}
m \frac{d^2y(t)}{dt^2}=-\bar{\gamma} \int_0^t(t-\tau)^{2H-2}
\frac{dy}{d\tau} d\tau +\eta\cdot\xi(t),
 \label{eq:thirty-two}
\end{equation}
with $d y(0)/dt=v_0$, $y(0)=0$, where $v_0$ is the initial velocity.
The solution of the stochastic Eq. (\ref{eq:thirty-two}) is
$$
\begin{array}{lll}
y(t)&=& v_0\cdot t\cdot E_{2H,2}(-\gamma \cdot
t^{2H})
\\
&&+\frac{\eta}{m} \int_0^t (t-\tau) \cdot E_{2H,2}(-\gamma \cdot
(t-\tau)^{2H})\xi(\tau)d\tau,
\end{array}
$$
\vskip 0.05cm \noindent where $\gamma=(\bar{\gamma}\Gamma(2H-1))/m$
and the generalized Mittag-Leffler function is
$$
E_{\alpha,\beta}(t)=\sum\limits_{n=1}^{\infty}
\frac{t^n}{\Gamma(\alpha n+\beta)},
$$
and $E_{\alpha,\beta}(-t) \thicksim (t \Gamma(\beta-\alpha))^{-1}$
when $t \rightarrow +\infty$.

We have
\begin{equation}
\langle y(t) \rangle=v_0\cdot t\cdot E_{2H,2}(-\gamma \cdot
t^{2H})\thicksim \frac{v_0}{\gamma}\frac{t^{1-2H}}{\Gamma(2-2H)} ,
\label{eq:thirty-three}
\end{equation}
and
\begin{equation}
\begin{array}{lll}
\langle y^2(t) \rangle &=& \frac{2k_BT}{m}t^2E_{2H,3}(-\gamma \cdot
t^{2H}) \\
\\
  &\thicksim&
\frac{2k_B T}{\bar{\gamma} \Gamma(2H-1)\Gamma(3-2H)}\cdot t^{2-2H},
\end{array}
 \label{eq:thirty-four}
\end{equation}
where the thermal initial condition: $v_0^2=k_B T /m$ is assumed.

Note that for short times we have $\langle y^2 (t) \rangle\sim (k_B
T /m)t^2$. Eqs. (\ref{eq:thirty-three}) and (\ref{eq:thirty-four})
were found
 \cite{Lutz:01,BarkaiSilbey:00}.

The covariance function of $y(t)$ reads
\begin{widetext}
\begin{equation}
\begin{array}{lll} \langle y(t_1) y(t_2)\rangle &=& v_0^2t_1t_2E_{2H,2}(-\gamma
t_1^{2H})E_{2H,2}(-\gamma t_2^{2H}) \\
\\
&& \displaystyle +\frac{k_BT\bar{\gamma}}{m^2}
\int_0^{t_2}\int_0^{t_1} d \tau d s \cdot (t_1-\tau)
E_{2H,2}(-\gamma(t_1-\tau)^{2H})(t_2-s)
E_{2H,2}(-\gamma(t_2-s)^{2H})|\tau-s|^{2H-2}.
\end{array}
 \label{eq:thirty-five}
\end{equation}
When $t_1,\,t_2$ tend to infinity,
\begin{equation} \langle y(t_1) y(t_2)\rangle \thicksim
\frac{k_BT}{\bar{\gamma}\Gamma^2(2H-1)\Gamma^2(2-2H)}\int_0^{t_2}
\int_0^{t_1} d \tau d s \cdot
(t_1-\tau)^{1-2H}(t_2-s)^{1-2H}|\tau-s|^{2H-2},\label{eq:thirty-six}
\end{equation}
\end{widetext}
i.e., the covariance of $y(t)$ approximates to the ones of $z(t)$,
so we can expect in the long-time limit

\begin{figure}[!htb]
\begin{center}
\includegraphics[width=9cm,height=8.5cm]{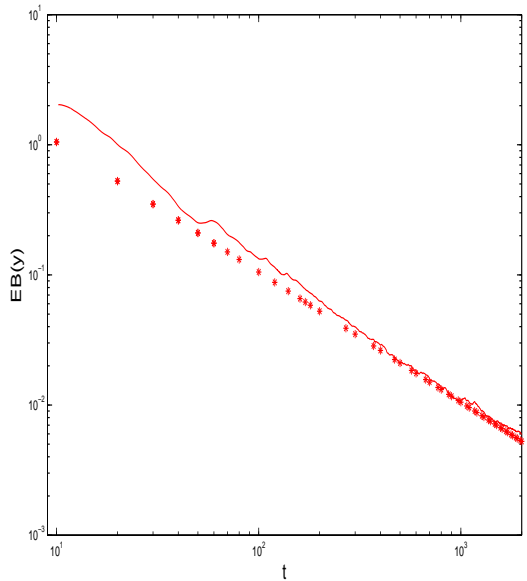}
\caption{\label{fig5} The ergodicity breaking parameter ${\rm
EB}(y)$ versus $t$. Simulations of $200$ trajectories were used with
$2-2H=0.75,~\Delta=10,~k_B=1,~m=1,~\bar{\gamma}=1,~v_0=1,~T=1$.  The
stars $*$ is the theoretical  result Eq.
(\ref{eq:MainResult}) with corresponding parameter values (without fitting).
}
\end{center}
\end{figure}
\begin{figure}[!htb]
\begin{center}
\includegraphics[width=9cm,height=8.5cm]{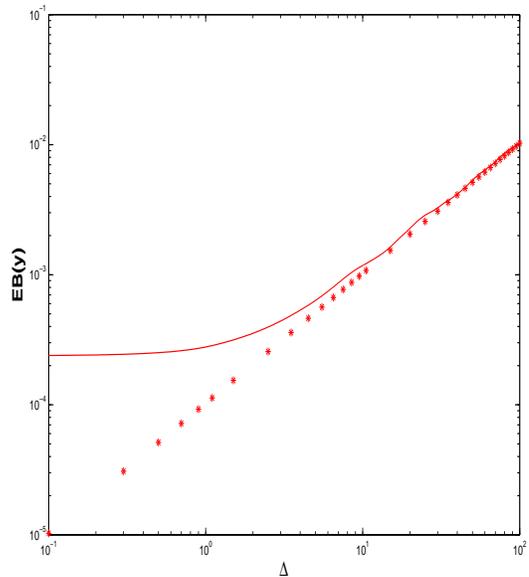}
\caption{\label{fig6} The ergodicity breaking parameter ${\rm
EB}(y)$ versus $\Delta$. Total 200 trajectories are used to compute
the average and variance, the measurement time $t=10^4$,
$2-2H=0.75,~k_B=1,~m=1,~\bar{\gamma}=1,~v_0=1,~T=1$. The stars $*$
are  the theoretical result Eq. (\ref{eq:MainResult}) with
corresponding parameter values. We see that results found from the
under-damped Langevin equation converge to our analytical theory
based on fBM.}
\end{center}
\end{figure}


\begin{equation}
\langle\overline{\delta^2}(y)\rangle \thicksim
\langle\overline{\delta^2}(z)\rangle\thicksim
\langle\overline{\delta^2}(x)\rangle,\label{eq:thirty-seven}
\end{equation}
and
\begin{equation}
{\rm EB}(y) \thicksim {\rm EB} (z) \thicksim {\rm EB} (x).
\label{eq:thirty-eight}
\end{equation}
The simulations \cite{Numerical:08}, Fig. \ref{fig4}, confirms Eq.
(\ref{eq:thirty-seven}) and Figs. \ref{fig5} and \ref{fig6}, support
Eq. (\ref{eq:thirty-eight}). Note that for short times we have a
ballistic behavior for $y(t)$ (see Fig. 4), but not for $z(t)$ and
$x(t)$, so clearly both $\Delta$ and $t$ must be large for Eq.
(\ref{eq:thirty-eight}) to
 hold.

\section{Discussion}

We showed that the fractional processes $x(t)$, $y(t)$ and $z(t)$
are ergodic. The ergodicity breaking parameter decays as a power-law
to zero. In the ballistic limit $\langle x^2 \rangle\sim t^2 $ non
ergodicity is found. For the opposite localization limit $\langle
x^2 \rangle \sim t^0$ (i.e.
 $H \to 0$ for fBM) the asymptotic convergence is reached only after
 very long times.  Our most surprising result is that the
transition between the localization limit and ballistic limit is not
smooth. When $H=3/4$ the EB is changed and the amplitude $k(H)$
diverges. Other very different critical exponents of fractional
Langevin equations were recently found in \cite{Burov:08}.  There
the critical exponents mark transitions between over-damped and
under-damped motion. So stochastic  fractional processes possess a
zoo of critical exponents.
 Let us now compare between our results and those derived
based on the CTRW model \cite{He:08}. The most striking difference
is that for the CTRW model we have non ergodic behavior even in the
long time limit.  It is attempting to conclude that this indicates
that the underlying stochastic motion for the mentioned experiments
in the cell is of CTRW nature. However, as mentioned in the
introduction experiments are conducted for finite times, and hence
what may seem as a deviation from ergodic behavior may actually be a
finite time effect. Here we gave analytical predictions for the
deviations from ergodicity for finite time measurement, based on
three fractional models. The EB parameter depends on measurement
time and lag time, and can be used to compare experimental data with
predictions of fractional equations (the EB parameter for the CTRW
is given in \cite{He:08}). It should be noted however that for
 sub-diffusion  in the
cell, effects of the boundary of the cell, may be important, and
these effects where not considered in this text. Another important
difference is that {\em for an infinite system} we have for the CTRW
$\langle \overline{\delta^2} \rangle \sim \Delta/ t^\alpha$, so the
time average procedure yields a linear dependence on $\Delta$ and an
aging effect with respect to the measurement time. Hence for CTRW an
anomalous diffusion process may seem normal with respect to $\Delta$
 \cite{He:08,Barkai:03a,Barkai:03b,Klafter:08}.
 In contrast for the fractional models we investigated
here, we have $\langle \overline{\delta^2} \rangle \sim
\Delta^\alpha$ which is the same as the ensemble average $\langle
x^2 \rangle \sim t^\alpha$. The main difference between the two
approaches, is that the CTRW process is non-stationary. It would be
interesting to
 investigate fractional Riemann-Liouville  Brownian
motion (Eq. (\ref{eq:one}) without the integral from $-\infty$ to
$0$) which is a non-stationary process.

{\bf Acknowledgement} This work was supported by the Israel Science
Foundation. EB thanks S. Burov for discussions.


\newpage 


\begin{thebibliography}{10}
\bibitem{Numerical:08} The detailed numerical
scheme and error analysis will be presented in the coming
publication.
\end{thebibliography}
\end{document}